\documentstyle[prd,aps,preprint,epsfig]{revtex}

\newcommand{\gev}{ {\rm GeV} }

\newcommand{\K}{ {\rm K} }
\newcommand{\kakko}[1]{{\left({#1}\right)}}

\newcommand{\mgrav}{ {m_{3/2}} }

\newcommand{\mev}{ {\rm MeV} }
\newcommand{\mG}{ {M_{\rm G}} }
\newcommand{\mphi}{ {m_{\phi}} }

\newcommand{\order}{{\cal O}}

\newcommand{\tev}{ {\rm TeV} }

\newcommand{\bea}{\begin{eqnarray}}  \newcommand{\eea}{\end{eqnarray}}
\newcommand{\beq}{\begin{equation}}  \newcommand{\eeq}{\end{equation}}
\newcommand{\bef}{\begin{figure}}  \newcommand{\eef}{\end{figure}}
\newcommand{\bec}{\begin{center}}  \newcommand{\eec}{\end{center}}
  
\newcommand{\lmk}{\left(}  \newcommand{\rmk}{\right)}

\begin{document}
\tighten

\preprint{
\noindent
\begin{minipage}[t]{3in}
\begin{flushright}
OU-TAP-241\\
TU-738 \\
hep-ph/0502211 \\
\end{flushright}
\end{minipage}
}

\title{Neutralino Dark Matter from Heavy Gravitino Decay}
\author{Kazunori~Kohri$^{1,2}$, Masahiro~Yamaguchi$^3$, and Jun'ichi~
Yokoyama$^2$}

\address{$^1$Harvard-Smithsonian Center for Astrophysics, 60 Garden
Street, Cambridge, MA 02138, USA}

\address{$^2$Department of Earth and Space Science, Graduate School of
Science,\\ Osaka University,
Toyonaka  560-0043, Japan}

\address{$^3$Department of Physics, Tohoku University, Sendai
980-8578, Japan}
\date{\today}

\maketitle
\begin{abstract}
    We propose a new scenario of non-thermal production of neutralino
    cold dark matter, in which the overproduction problem of lightest
    supersymmetric particles (LSPs) in the standard thermal history is
    naturally solved.  The mechanism requires a heavy modulus field
    which  decays mainly to ordinary particles releasing large entropy
    to dilute  gravitinos produced just after inflation and thermal
    relics of LSPs.  Significant amount of gravitinos are also
    pair-produced at the decay, which subsequently decay into the
    neutralinos. We identify the regions of the parameter space in
    which the requisite abundance of the neutralino dark matter is
    obtained without spoiling the big-bang nucleosynthesis by
    injection of hadronic showers from gravitino decay.  The
    neutralino abundance obtained in this mechanism is insensitive to
    the details of the superparticle mass spectrum, unlike the
    standard thermal abundance. We also briefly mention the
    testability of the scenario in future experiments.

\end{abstract}




\section{Introduction}

Supersymmetry (SUSY) has been a promising candidate for new 
physics beyond
the standard model, not only because it is an attractive solution
to the naturalness problem associated with the electroweak scale, but
also because it provides a natural candidate for cold dark matter. 
Indeed, the  relic abundance of the lightest supersymmetric particle (LSP),
which is stable by virtue of the R parity conservation,
may naturally fall into the critical density with plausible values of its
mass and couplings under the standard thermal history of the early universe.  
This concordance was regarded as one of the triumph of particle
cosmology in the last century.

Recent progress on the determination of the the cold dark matter (CDM)
abundance, however, has cast a shadow on the above simplistic view of
the standard SUSY cosmology.  In particular, the value indicated by
the WMAP measurement pushes the preferred regions of the SUSY
parameter to only tiny and special corners of the parameter space.  In
fact the WMAP data implies the Dark Matter
abundance~\cite{Spergel:2003cb}
\begin{eqnarray}
    \Omega_{\rm{CDM}} h^2 = 0.1126^{+0.0161}_{-0.0181}
\end{eqnarray}
at  95 $\%$  C.L.,  where $\Omega_{\rm{CDM}}$  represents the  current
value of  the density parameter of  the CDM component, and  $h$ is the
Hubble parameter in  unit of 100 km/s/Mpc.  This  is much smaller than
unity. On the other hand, assuming the standard thermal history of the
universe,  the  thermal relic  abundance  of  the  neutralinos can  be
computed in a standard manner. Importantly it is known that in generic
regions of the  SUSY parameter space, the relic  abundance computed in
this way tends to exceed the value the WMAP
suggests~\cite{Ellis:2003cw}.  The right amount of  the neutralino
abundance is  obtained  only in  tiny and  special regions of  the
parameter space  where some enhancement  mechanisms of neutralino
annihilation take place. They include 1) light sfermions, 2)
coannihilation  with  e.g.  chargino,  stau,  or  stop,  3)  resonance
enhancement,  and 4)  annihilation into  W pair.

Furthermore this tension we now have may become real conflict if the
LHC experiments, for instance, will find SUSY particles outside the
regions where the calculated relic abundance is consistent with that
implied by the WMAP measurement.

At this point, one should recognize that the argument of the thermal
relic abundance strongly relies on the standard thermal history of the 
universe.  The fact that the enhanced annihilation of neutralinos is
 required to meet observation may imply that the universe might have
 evolved in a non-standard manner.  Alternatively, we may find a natural
 solution to the dark matter problem making use of a non-standard
 thermal history without any enhancement of the annihilation processes, 
which is the central topic of the present paper.

Tension in the SUSY cosmology with the standard thermal history
increases even more when one turns to the recent reanalysis of the
gravitino problem which includes the effects of hadronic showers on
the abundance of the light elements produced by big-bang
nucleosynthesis (BBN) \cite{Kawasaki:2004qu}.  It is well-known that
decays of long-lived particles such as gravitinos would spoil the
success of the big-bang nucleosynthesis because of the production and
destruction of the light elements by emitted high-energy photons or
hadrons~\cite{BBNwX_OLD,Reno:1987qw,DimEsmHalSta,KawMor,photodis_new}.
If the gravitino mass is just around the weak scale, very severe upper
bounds come from the overproduction of the $^6$Li abundance and the D
abundance. We should also stress that inclusion of the hadronic shower
also put a rather stringent bound for heavier gravitino whose lifetime
is even shorter than 1 sec  This is because the hadronic shower
produces pions and nucleons which interact with ambient nucleons in
the thermal bath, changing the neutron to proton ratio and hence the
$^4$He abundance~\cite{Reno:1987qw,Kohri:2001jx}.  Note that the
argument is based on the standard thermal history of the universe in
which there is no dilution or extra production of gravitinos after
reheating epoch of primordial inflation \cite{Sato:1980yn,lindebook}.

It is, however, quite plausible that the universe did not follow the
standard thermal history described above. This is particularly the
case when there are some moduli fields whose masses are not far away
from the electroweak scale.  The coherent oscillation of the moduli
fields is likely to dominate the energy density of the universe at
some epoch in the early universe, followed by rather late decays
\cite{Coughlan:1983ci}.  If the mass is in the electroweak scale, the
lifetime of the moduli fields is much longer than 1 sec, meaning that
the entropy production at the moduli decays takes place with reheat
temperature lower than 1 MeV. This would upset the success of the
BBN. The disaster can be avoided if the moduli masses are somewhat
larger than the electroweak
scale~\cite{Moroi:1994rs,Randall:1994fr,Kawasaki:1999na,Hannestad:2004px}.

In a previous paper \cite{Kohri:2004qu}, we have shown that the
gravitino abundance can be regulated to an observationally acceptable
level if we take entropy production due to moduli decay into account.
In fact we have shown that it is possible to satisfy 
the severe constraints imposed by its hadronic decay \cite{Kawasaki:2004qu}
even when the reheat temperature after inflation was as high as $\sim
10^{16}$GeV and the new production mechanism of gravitinos from moduli
decay is taken into account \cite{Kohri:2004qu}.

In this paper, we would like to explore this line of argument and
propose a new scenario in which the neutralino dark matter is produced
in a non-thermal way. In our scenario, the neutralinos of the thermal
origin are all diluted by the entropy production at the moduli
decay. The dark matter neutralinos are produced, however, by the decay
of gravitinos, following the moduli decay to gravitinos.
 
It is very non-trivial whether this mechanism survives severe
constraints from BBN on the gravitino decay which produces hadronic
and/or electromagnetic showers. We will show that in a range of the
mass parameters, the gravitino decay survives the constraint from the
BBN, whereas the abundance of the neutralino LSPs is in accord with
the cosmological observations.  Thus our model offers a consistent
scenario of cosmic evolution in the new era of precision cosmology
where the cold dark matter abundance is much smaller than previously
expected and the gravitino problem is much more aggravated.  We will
also see that the scenario requires small hierarchy among moduli,
gravitino and soft SUSY breaking masses.  We will argue how this small
hierarchy is naturally realized.

The organization of the paper is as follows.  In the next section we
outline the basic features of our scenario.
In \S III we describe consequences of moduli decay, namely, amount of
entropy and gravitinos from them.  Neutralino abundance from decaying
gravitino is studied in \S IV.  Then we compare our yield with the
constraints imposed by BBN in \S V.  Finally \S VI is devoted to
conclusion and discussion.

\section{A successful scenario with small hierarchy: Our proposal}

In superstring theory and supergravity, there often exist scalar
fields which have (almost) flat potential.  Examples include moduli
fields and the dilaton field in classical string vacua, whose vacuum
expectation values (VEVs) determine the size and shape of compactified
extra dimensions and the string coupling constant, respectively.  They
also include the field which is responsible for supersymmetry breaking
in a typical gravity mediated supersymmetry breaking scenario (the
Polonyi field).  They typically have Planck-scale suppressed
interaction, and the VEVs of the order of the Planck scale.  We will
generically call these fields the moduli fields.

It is expected that these moduli fields acquire non-flat potential and
thus masses in some manner.  Their masses depend on the mechanism how
the moduli are stabilized.  In a conventional view~\cite{Coughlan:1983ci}, 
the moduli fields
acquire superpotential from some effect associated with
supersymmetry breaking. In this case one naively expects that their
masses are of the order of the gravitino mass, a typical mass scale of
supersymmetry breaking in supergravity.

Recent development of flux compactification reveals the possibility
that moduli yield superpotential when higher-form antisymmetric tensor
fields acquire VEVs in compact directions~\cite{flux}. The flux
compactification fixes some of the moduli fields, but not all. In
particular, in the type II superstring the overall modulus whose VEV
determines the size of the compactified manifold is not stabilized in
this way.  These remaining moduli may acquire potential by other
non-perturbative effects, and we may expect that their masses are
related to the gravitino mass scale.\cite{Kachru:2003aw,Choi:2004sx}

Thus the naive expectation of the gravity mediation of supersymmetry
breaking is that the masses of the moduli fields and sparticles in the
MSSM sector are similar, and is of the order of the gravitino
mass. Since the sparticle masses are expected to be in (sub-)TeV
range, the naive guess leads that the gravitino mass as well as the
moduli masses are also around this range.  In fact, the cosmological
moduli problem and the gravitino problem are consequences of this
naive expectation.

A closer look at these masses, however, reveals possible deviation of
the mass scales in one or two orders of magnitude.  In the original
gravity mediation model by Polonyi, the SUSY mass scale is given by
hand, and the mass of the Polonyi field is indeed comparable to the
gravitino mass. In more realistic models, the potential of the moduli
fields may be generated dynamically. A typical example of this comes
from gauge dynamics such as gaugino
condensation~\cite{gaugino-condensate}.  In this scheme, the
hierarchically small SUSY breaking scale is achieved by the
dimensional transmutation with some numerical coefficient of order
10-100. At the same time, the moduli mass which is determined by the
second derivative of the scalar potential can be larger by the same
amount than the gravitino mass which is naively expected. In fact,
this is the case for the well-known racetrack
scenario~\cite{Krasnikov:1987jj} where multiple gaugino condensates
take place to obtain a non-trivial minimum of the scalar potential.
Similar observation has been made in other scenarios, including
non-perturbative correction to the K\"{a}hler potential of the dilaton
field~\cite{Banks:1994sg}.  For explicit demonstration,
see~\cite{Buchmuller:2004xr}.  
The moduli masses can also be heavy in some scenarios of flux 
compactification~\cite{Choi:2004sx}.

This consideration sets a small hierarchy between the moduli masses
and the gravitino mass. Another small hierarchy between the gravitino
mass and the sparticle masses in the MSSM can be realized when the
field responsible for the SUSY breaking is somewhat sequestered from
the MSSM sector~\cite{Inoue:1991rk,Randall:1998uk,Luty:1999cz}. 
It is interesting to note that the coupling of the
moduli fields to the MSSM sector does not violate this property of
sequestering, because the moduli fields are heavy so that their
auxiliary fields have SUSY breaking VEVs suppressed by the moduli
masses relative to the gravitino mass.  Then the softly SUSY breaking
masses for sparticles are a combination of the contribution from the
moduli SUSY breaking and that from the anomaly 
mediation.\cite{Randall:1998uk,Giudice:1998xp}

The small mass hierarchy among the  moduli, gravitino and sparticle masses
given above yields a successful scenario without the difficulties of 
the gravitino as well as the neutralino abundance. Here we would like
to sketch our proposal. 
From now on, we assume, for simplicity, that there is only one modulus field
which is relevant in our discussion.

The heavy modulus field is expected to obey damped coherent oscillation
with the initial amplitude of order the Planck scale.  The modulus
oscillation dominates the energy density of the universe until it
decays to reheat the universe. In the present case, since the modulus
is rather heavy, its lifetime can be much shorter than 1 sec. The
entropy production at the reheating dilutes the primordial abundances
of the gravitinos. Non-negligible amount of gravitinos are, however,
produced by the modulus decay, on which  the constraint from BBN will be
imposed.

The sparticles in the MSSM sector are also diluted by the modulus
decay, if the decay products do not contain the sparticles and the
thermal bath after the reheating is not hot enough to regenerate the
sparticles by scattering processes. We will check that the former is
indeed the case when the modulus coupling to the MSSM sector is of certain
type, and the latter will be guaranteed if the reheating temperature
is low enough. In this case, the neutralinos which remain until today
come from the gravitino decay. The main assertion of this paper is that in
some regions of the parameter space the neutralinos can be abundant to
constitute the dark matter of the universe, whereas the gravitino
abundance satisfies the constraint from the BBN.
 
The rest of the paper is devoted to describe the details of our proposal
given above.

\section{Moduli Decay: Entropy Production and Gravitino Production}
The modulus field is typically displaced from the true minimum during
the inflation, which starts damped coherent oscillation when the
expansion rate of the universe decreases to the modulus mass. Unless
 the initial amplitude of the modulus field $\phi_i$ is  very small
compared to the Planck scale, the energy density of the modulus field
in the form of the coherent oscillation will soon dominate the
universe.  The modulus eventually decays.  The decay depends on how it
couples to other particles \cite{Moroi:1999zb}.  Here we assume that it
decays mainly to the ordinary particles in the MSSM and thus the
universe becomes radiation dominated again. This is especially the
case when the modulus coupling to the gauge bosons and gauginos
through gauge kinetic function gives the largest contribution to the
decay, where the modulus decays dominantly to a gauge boson pair while
the decay to a gaugino pair suffers chirality suppression.  We 
write the total decay width of the modulus field
\begin{eqnarray}
    \Gamma_{\rm{tot}}=N \frac{m_{\phi}^3}{M_G^2},
\end{eqnarray}
where $m_{\phi}$ is the mass of the modulus field, $M_G=2.4\times
 10^{18}$
GeV represents
the reduced Planck scale, and $N$ parameterizes our ignorance of the
interaction of the modulus. It is natural to expect that $N\sim 10^{-2}-1$.
  Given the above expression, the
reheat temperature after the modulus decay $T_R$ is calculated to
be 
\begin{eqnarray}
\label{eq:TR}
     T_R&= &\left( \frac{90}{\pi^2 g_*} \right)^{1/4} 
           \sqrt{\Gamma_{\rm{tot}}M_G}
        =\left( \frac{90 N^2}{\pi^2 g_*} \right)^{1/4}
             m_{\phi}^{3/2} M_G^{-1/2} 
\nonumber \\ 
       &\approx & 11 N^{1/2} \left(\frac{g_*}{10^2}\right)^{-1/4}
                          \left( \frac{m_{\phi}}{10^4 \rm{TeV}}\right)^{3/2}
                          \rm{GeV},
\end{eqnarray}
where $g_*$ counts the effective relativistic degrees of freedom with 
$g_*=228.75$ for the MSSM particles. 

Entropy production is also computed in a straightforward way. 
The entropy increase factor is found to be \cite{Kohri:2004qu}
\begin{eqnarray}
          \Delta=\sqrt{\frac{2}{9N}}\frac{M_G}{m_{\phi}}
                   \left(\frac{\phi_i}{M_G}\right)^2
\end{eqnarray}
Thus, as far as the modulus mass is much lower than the Planck scale
and the initial amplitude of the modulus oscillation $\phi_i$ is not
extremely small, the entropy increase factor is much larger than
unity, which we assume to be the case. The gravitino abundance as well
as the sparticle abundances before the reheating will be diluted.

Let us now discuss the production of the gravitinos by the modulus decay.
It has been pointed out in \cite{Hashimoto:1998mu} 
that the modulus field generically decays to
a pair of the gravitinos via the following terms in the supergravity
Lagrangian:
\begin{eqnarray}
       {\cal L}_{\rm int}=-e^{K/2} W \bar{\psi}_{\mu} \sigma^{\mu \nu}
                                  \bar{\psi}_{\nu} +H.c.
\end{eqnarray}
where $\psi_{\mu}$ represents the gravitino field, $K$ is the K\"ahler 
potential, and $W$ is the superpotential. In the above expression, we
have  the reduced Planck scale $M_G$ to be unity.  Since $K$ and $W$ have
the moduli dependence, the above term in the Lagrangian gives the coupling
of the modulus to the gravitino pair.  Explicitly this is given by
\begin{eqnarray}
    \left( e^{K/2} W \right)_{\phi}=
   \frac{K_{\phi}}{2} e^{K/2} W + e^{K/2} W_{\phi},
\label{eq:modulus-gravitino}
\end{eqnarray}
where the suffix $\phi$ here stands for the derivative with respect 
to $\phi$.  Now the auxiliary component of the modulus field is written
\begin{eqnarray}
  \bar{F}^{\bar{\phi}}=-e^K (K^{-1})^{\bar{\phi} \phi}
             \left( W_{\phi} +K_{\phi} W \right).
\end{eqnarray}
As we assume that the modulus field is not a dominant source of SUSY breaking, 
we can approximately equate $W_{\phi}=-K_{\phi} W$. Using this,
(\ref{eq:modulus-gravitino}) can be rewritten as follows:
\begin{eqnarray}
       \left( e^{K/2} W \right)_{\phi}
    = -\frac{1}{2}K_{\phi} e^{K/2} W = -\frac{1}{2}K_{\phi} m_{3/2}.
\end{eqnarray}
Thus  the partial decay width to the gravitino pair becomes of the form
\begin{eqnarray}
   \Gamma (\phi \rightarrow 2 \psi_{\mu})=
   C \frac{m_{3/2}^2}{M_G^2}m_{\phi}=C \frac{m_{\phi}^3}{M_G^2}
     \left( \frac{m_{3/2}}{m_{\phi}} \right)^2,
\end{eqnarray}
where we have recovered the reduced Planck scale $M_G$ explicitly.
In this expression, we have introduced a parameter 
$C\sim 1/32\pi |K_{\phi}/M_G|^2$, where a possible numerical factor as well as
the phase space factor is ignored. 
 We expect that $C\sim 10^{-2}-10^{-1}$
would be natural, but a smaller $C$ may also be possible.

It has recently been shown that a more effective decay channel to the
gravitino and the modulino, the fermionic component of the modulus
supermultiplet, is kinematically allowed if a certain coupling of the
modulus to the field  responsible to the SUSY breaking is large enough
\cite{Kohri:2004qu}.  Whether this channel is allowed or not is,
however, quite model dependent.  Here we conservatively consider the
former decay mode to the gravitino pair, which will also provide a
preferred result as we will see shortly.

Gravitinos are also produced by scattering processes in the thermal
bath after the reheating epoch.  The production rate is proportional
to the reheat temperature and the ratio of the number density of
thermally produced gravitinos to the entropy density after modulus
decay reads~\cite{Moroi:1995fs,Bolz:2000fu,Kawasaki:2004qu},
\bea
\label{eq:thermal}
Y_{3/2}^{\rm th}&\simeq& 1.6 \times 10^{-12} 
             \left(\frac{T_{R}}{10^{10} \gev}\right) \nonumber \\
     &=& 2 \times 10^{-9}\lmk\frac{N}{10}\rmk^{1/2}
\lmk\frac{g_*}{10^2}\rmk^{-1/4}\lmk\frac{\mphi}{10^5 \rm
             TeV}\rmk^{3/2}.
\eea
With the reheat temperature much below 1 TeV which we are interested
in, the gravitino production by the scattering processes is negligibly
small.

As was shown in \cite{Kohri:2004qu}, the gravitino problem  will be
solved in this scenario. The primordial abundance of the gravitino
generated before the reheating due to the modulus decay will be
diluted to a negligible level.  Although the gravitinos are produced
by the modulus decay, the abundance was shown to be small enough to
survive the constraints from the big-bang nucleosynthesis in an
appropriate region of the parameter space. Here, we consider the decay
mode of the modulus to the gravitino pair. Then the allowed region in
the parameter space will become larger than that of
Ref.~\cite{Kohri:2004qu}.

More explicitly, let us estimate the gravitino abundance  produced by
the modulus decay.  Here we approximate the decay of the  modulus
field as an instantaneous decay. The energy density of the modulus
field is equal to  that of the radiation at the reheating, {\em i.e.}
at  the modulus decay:
\begin{eqnarray}
      \rho_R=m_{\phi} n_{\phi}
\end{eqnarray}
where $n_{\phi}$ is the number density of the modulus. The gravitino number
density at the modulus decay $n_{3/2}$ 
can also be easily estimated in the approximation
\begin{eqnarray}
   n_{3/2}=2 B n_{\phi}
\end{eqnarray}
where $B$ represents the branching ratio of the modulus decay into the
gravitino pair
\begin{eqnarray}
     B\equiv {\rm Br}(\phi \rightarrow 2 \psi_{\mu})
      =\frac{C}{N}\left( \frac{m_{3/2}}{m_{\phi}} \right)^2.
\end{eqnarray}
The gravitino abundance in the form of yield $Y_{3/2}$, 
the ratio of the number density relative to the entropy density, is
given by
\begin{eqnarray}
\label{eq:direct}
     Y_{3/2}=\frac{3}{4}\frac{n_{3/2}T_R}{\rho_R}
            =\frac{3}{2} B \frac{T_R}{m_{\phi}} 
            =1.8\times 10^{-10}CN^{-1/2}
            \lmk\frac{g_*}{10^2}\rmk^{-1/4}\lmk\frac{\mgrav}{10^2 \rm
            TeV}\rmk^2 \lmk\frac{\mphi}{10^5 \rm TeV}\rmk^{-3/2}.
\end{eqnarray}

\section{Neutralino Abundance}
In this section, we would like to discuss the relic abundance of the
neutralinos. Here we assume that the R-parity is conserved and  the
lightest neutralino becomes the lightest superparticle, and thus it is
a candidate for dark matter of the universe.

The entropy production at the modulus decay completely dilutes the
primordial abundance of the neutralinos and other superparticles.
Thus in our scenario the neutralinos which remain today should be 
produced at and after the reheating due to the modulus decay. There
are three possible sources of the neutralinos:
\begin{itemize}
\item Neutralinos produced by gravitino decay
\item Neutralinos produced by modulus decay
\item Neutralinos produced in the thermal bath during and after the
reheating
\end{itemize}
We now show that the first contribution can  dominate over the other two.

As we discussed earlier, the universe is reheated at the modulus
decay. If the reheat temperature were too high, the neutralinos would
again be in the thermal equilibrium. In this case the standard
computation of the neutralino relic abundance would apply, resulting
in the potential problem of the too large  relic abundance unless some
enhancement mechanism of their annihilation is at work. This would be
the case if the reheat temperature exceeded $m_{\chi}/20 \sim
m_{\chi}/25$, where $m_{\chi}$ is the mass of the lightest neutralino.
On the other hand, if the reheat temperature is fairly low, the
regeneration of the neutralinos in the thermal bath is negligibly
small.  Therefore we will consider the case where $T_R <m_{\chi}/30$
in the following. Note that this condition is not the constraint, but
the requisition in the current interest.

Let us next discuss the neutralinos produced by the gravitino decay.
The point here is that the gravitino decays eventually to
the neutralino LSP under the R-parity conservation, and the number
of the neutralinos is essentially the same as that of the gravitinos.
Thus the yield of the neutralino LSPs is evaluated as
\begin{eqnarray}
  Y_{\rm LSP} =Y_{3/2}=1.8\times 10^{-10}CN^{-1/2}
\lmk\frac{g_*}{10^2}\rmk^{-1/4}\lmk\frac{\mgrav}{10^2 \rm TeV}\rmk^2
\lmk\frac{\mphi}{10^5 \rm TeV}\rmk^{-3/2}. 
\label{eq:YLSP}
\end{eqnarray}
This is a constant of time, if there is no other process which will
change the number of the neutralinos nor the total entropy.  We would
like to emphasize that the neutralino abundance obtained in this
mechanism is insensitive to the details of the superparticle mass
spectrum, unlike the standard thermal abundance because $Y_{\rm LSP}$
is directly related with only $Y_{3/2}$.  Annihilation of the
neutralinos, if it took place, would change it. We will come back to
this issue later on. On the other hand, using the density parameter of
LSP, $Y_{\rm LSP}$ is also represented by
\begin{eqnarray}
  \label{eq:ychi_omega} 
  Y_{\rm LSP} = 3.65 \times 10^{-11} \Omega_{\rm
  LSP} h^2 \left(\frac{m_{\rm LSP}}{100 \gev} \right)^{-1}\left(
  \frac{T_0}{2.725 \K} \right)^{-3},
\end{eqnarray}
where we adopted the present photon temperature $T_0 = 2.725 \pm
0.002~\K$~\cite{mather:1999}.

Neutralinos may also be produced by the modulus decay. Let us again
consider the case where the modulus couples to the bilinear of the
supersymmetric form of the gauge potential. In this case, the decay to
the gaugino pair is known to receive chirality suppression, and thus
the branching ratio is roughly estimated as
\begin{eqnarray}
   {\rm Br (\phi \rightarrow 2 \tilde{g}) }
  \sim \left( \frac{m_{\tilde{g}}}{m_{\phi}} \right)^2
\end{eqnarray}
where $\tilde g$ symbolically denotes a gaugino. This should be compared
with the branching ratio to the gravitino pair as
\begin{eqnarray}
\frac{ \rm{Br}(\phi \rightarrow 2 \tilde{g})}
{ \rm{Br}(\phi \rightarrow 2 \psi_{\mu})}
\sim \frac{N}{C} \left( \frac{m_{\tilde{g}}}{m_{3/2}} \right)^2.
\end{eqnarray}
Note that we are considering the case where $(m_{\tilde{g}}/m_{3/2})^2
\sim 10^{-4} -10^{-3}$. Thus the contribution to the neutralino
abundance from the direct modulus decay is sub-dominant, as far as the
ratio $N/C$ does not exceed $\sim \order(10^{4})$.

The addition of the coupling of the modulus to two chiral multiplets
({\em e.g.} quarks and their superpartners) does not change the
conclusion given above, because the decay to a squark pair, for
instance, is suppressed by the fourth power of the mass ratio
$m_{\tilde q}/m_{3/2}$, unlike the second power for the case of the
gaugino pair~\cite{Moroi:1999zb}. Thus it is negligible, provided that
the coupling of the modulus to the gauginos is not small. In the
following, we shall not consider the direct modulus decay to
superparticles in the MSSM.

In the following we assume that the pair annihilation of the
neutralinos after the production by the gravitino decay does not take
place. Whether the annihilation is effective or not can be seen by
comparing the annihilation rate to the Hubble expansion rate of the
universe when the neutralinos are produced by the gravitino decay.
Although the annihilation cross section of the neutralino $\sigma_{\rm
ann}$ is a function of the superparticle mass spectrum, it is
generically bounded as
\begin{eqnarray}
    \label{eq:sigma_ann}
    \sigma_{\rm ann} \lesssim  \frac{\alpha_{1}}{m_{\chi}^{2}},
\end{eqnarray}
with the coupling $\alpha_{1} \sim 10^{-2}$. Then, the ratio of the
annihilation rate $\Gamma_{\rm ann} = n_{\rm LSP}\sigma_{\rm ann}$, to
the Hubble expansion rate  $H \sim T^{2}/\mG$ is bounded as
\begin{eqnarray}
    \label{eq:gamma_to_H}
    \frac{\Gamma_{\rm ann}}{H} < \sigma_{\rm ann} \
    y_{3/2}^{\rm max} \ s \ \frac{\mG} {T^{2}} 
    = 10^{-4} \kakko{\frac{m_{\chi}} {100 \ \gev}}^{-3}
    \kakko{\frac{T}{\mev}}. 
\end{eqnarray}
Therefore, we find $\Gamma_{\rm ann}/H < 1$ at the temperature of
gravitino decay.

\section{Results}
We are ready to present our numerical results now.  We first discuss
constraints we impose on the parameter space.  Constraints on the
reheat temperature after moduli decay are that it should be 1) higher
than the lower bound required by the successful BBN, $T_R >1.2$
MeV~\cite{Kawasaki:1999na},~\footnote{Here we adopted a milder lower
bound on the reheat temperature when we assume that the modulus decays
only into photons. On the other hand, if we consider the case that the
decaying modulus also emits hadrons, the lower bound becomes higher
$T_R \gtrsim$ 4--5 MeV~\cite{Kawasaki:1999na}. However, this
difference does not change our results at all for the current purpose.}
but 2) lower than the temperature above which the neutralinos are
thermalized.  We set the latter condition as $T_R < m_{\rm LSP}/30$.

\begin{figure}[hp]
\begin{center}
\epsfxsize=0.9\textwidth\epsfbox{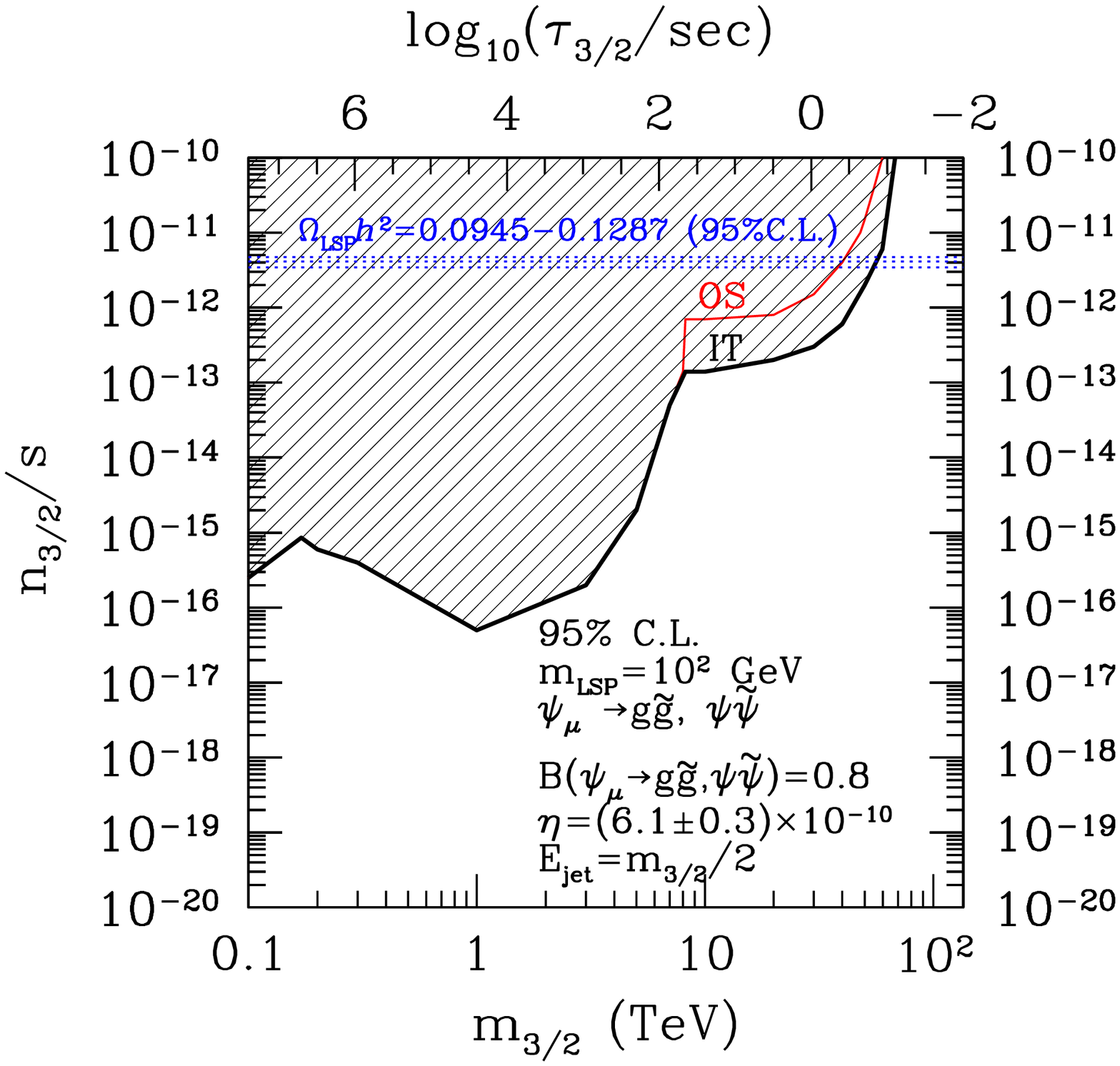}
\caption{ Upper bound on the yield variable $y_{3/2}=n_{3/2}/s$. For
heavy gravitinos with their masses of $m_{3/2} > 10 \tev$, the
constraint mainly comes from the $^4$He overproduction. The thick
solid line, which is denoted by ``IT'', comes from the observational
value of $^4$He mass fraction by Izotov and Thuan (2003). The dotted
strip indicates the region where the LSP abundance from the gravitino
decay corresponds to the desired value $\Omega_{\rm
LSP}h^2=0.0945-0.1287~(95\%~{\rm C.L.})$. The thin solid line, which
is denoted by ``OS'', is the case that we adopted the most
conservative observational value of $Y_p$ by Olive and Skillman
(2004). Here we assumed that a gravitino can decay into all of the
species of the chiral fermion and its scalar partner pairs
($\psi\tilde{\psi}$) and the gauge boson and the gaugino pairs
($g\tilde{g}$). }
\label{fig:m_n32_new}
\end{center}
\end{figure}

We show the constraint on gravitino decay from the BBN in
Fig.~\ref{fig:m_n32_new}. The upper bound on the yield of the
gravitino $n_{3/2}/s$ is drawn as a function of the gravitino
mass. Effects of hadronic shower are taken into account
\cite{Kawasaki:2004qu}.  Here we take the hadronic branching ratio to
be 0.8 because such a heavy gravitino can decay into all of the gauge
boson and the gaugino pairs and the chiral fermion and its scalar
partner pairs.~\footnote{According to the treatments in
Ref.~\cite{Kawasaki:2004qu}, we could not include all of the hadronic
decay modes into the BBN computation at one time. Therefore, here we
consider only the most dominant mode into the hadrons, i.e., the
gluon-gluino pairs. Then the branching ratio into the gluon-gluino
pairs is approximately 0.5. Because we adopted this value for the BBN
computation, the obtained bound on $y_{3/2}=n_{3/2}/s$ in Fig.~1 might
be milder by $\sim 30 \%$ approximately. Note, however, that this
difference of the hadronic branching ratio does not change the
constraints on the model parameters so much nor does our results at
all. For the more accurate computations, in which we include all of
the hadronic decay modes at one time, we would like to discuss them in
separate issues.}  For gravitino mass lighter than $10$ TeV, the
constraint comes from the overproduction of $^6$Li and D
abundances. On the other hand, the case with heavier mass is
constrained by the $^4$He overproduction. This is because the
inter-converting reaction between the neutron and the proton becomes
effective, which is induced by the pions and nucleons emitted by the
hadronic decays of the gravitinos. Here we adopted the observational
value of the $^4$He mass fraction $Y_p(IT) = 0.242 \pm$~0.002
(stat)~($\pm$~0.005 (syst)) by Izotov and Thuan
(2004)~\cite{Izotov:2003xn}.~\footnote{ Recently much milder value of
the observational $^4$He mass fraction was reported by Olive and
Skillman, $Y_p(OS) = 0.249 \pm$~0.009\cite{Olive:2004kq}. Because
their method is not the standard way to obtain the temperature and the
electron density, we adopt it as a reference value here. The
constraint which comes from $Y_p(OS)$ is denoted by the thin solid
line in the figures hereafter.}  In the same figure, the dotted strip
indicates the region where the LSP abundance from the gravitino decay
corresponds to the desired value $\Omega_{\rm
LSP}h^2=0.0945-0.1287~(95\%~{\rm C.L.})$ in Eq.~(\ref{eq:ychi_omega}).
Here we have assumed the LSP mass, $m_{\rm LSP}$, to be 100 GeV.  One
can easily scale the strip for other values of $m_{\rm LSP}$ by using
Eq.~(\ref{eq:ychi_omega}).

\begin{figure}[hp]
\begin{center}
\epsfxsize=0.9\textwidth\epsfbox{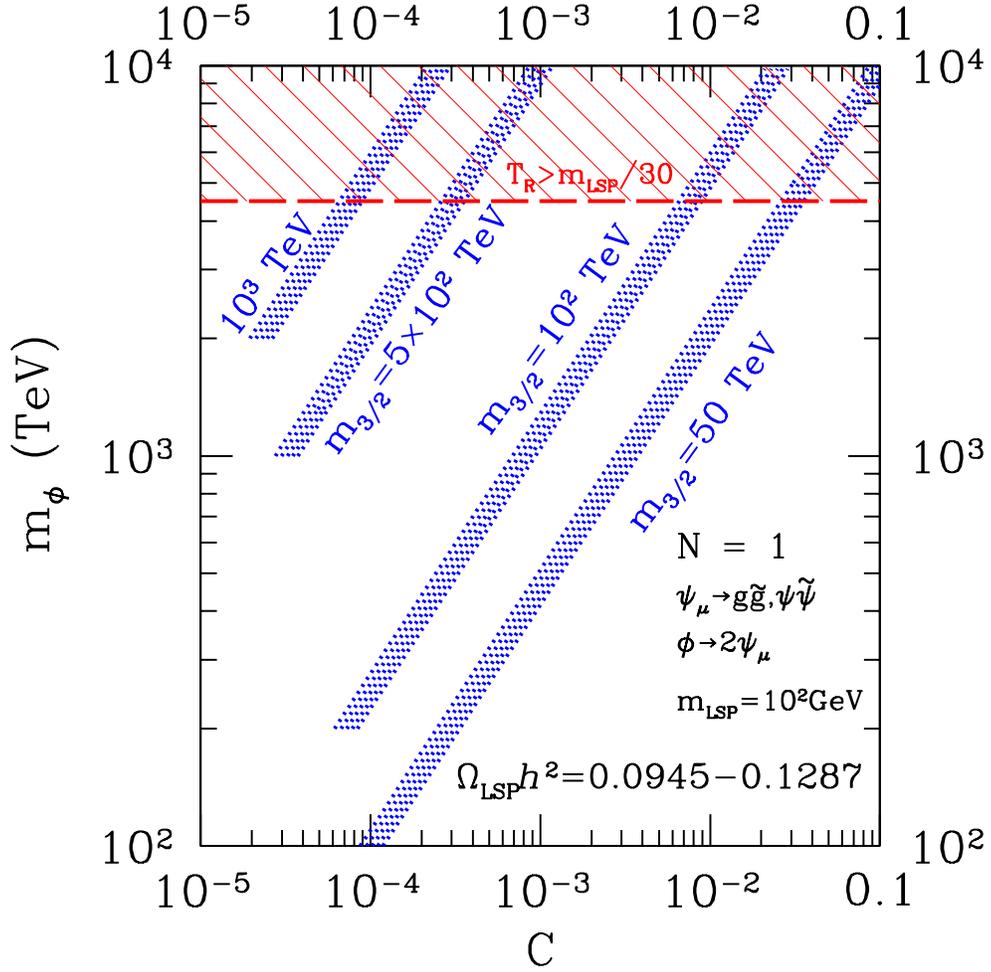}
\caption{Allowed regions for suitable observational value of
$\Omega_{\rm LSP}h^2=0.0945-0.1287~(95\%~{\rm C.L.})$ in the
$C$--$m_{\phi}$ plane. We plotted the strips for $m_{3/2}$ = 50,
$10^2$, $5\times 10^2$, and $10^3$ TeV, respectively. We considered
only two conditions that $T_R < m_{\rm LSP}/30$ (below the long dashed
line), and $m_{\phi} > 2 m_{3/2}$. Here we took $N = 1$, and $m_{\rm
LSP} = 100~\gev$.}
\label{fig:c_mphi}
\end{center}
\end{figure}

By equating  Eq.~(\ref{eq:YLSP}) with Eq.~(\ref{eq:ychi_omega}), we
see that the allowed value of $m_{\phi}$ is proportional to $C^{2/3}$
for a fixed value of $\Omega_{\rm LSP}h^2$.  In Fig.~\ref{fig:c_mphi},
we plot the allowed regions for suitable observational value of
$\Omega_{\rm LSP}h^2=0.0945-0.1287~(95\%~{\rm C.L.})$ in the
$C$--$m_{\phi}$ plane. We plotted the strips for $m_{3/2}$ = 50,
$10^2$, $5\times 10^2$, and $10^3$ TeV, respectively. We considered
only two conditions that $T_R < m_{\rm LSP}/30$ and $m_{\phi} > 2
m_{3/2}$.  Here we took $N = 1$, and $m_{\rm LSP} = 100~\gev$. As a
result we find that $C$ should lie approximately in the range of
$10^{-4}\lesssim C \lesssim 10^{-1}$.  Note that the theoretically
natural value of $C\sim 10^{-2}-10^{-1}$ is completely included in
this range.

\begin{figure}[hp]
\begin{center}
\epsfxsize=0.9\textwidth\epsfbox{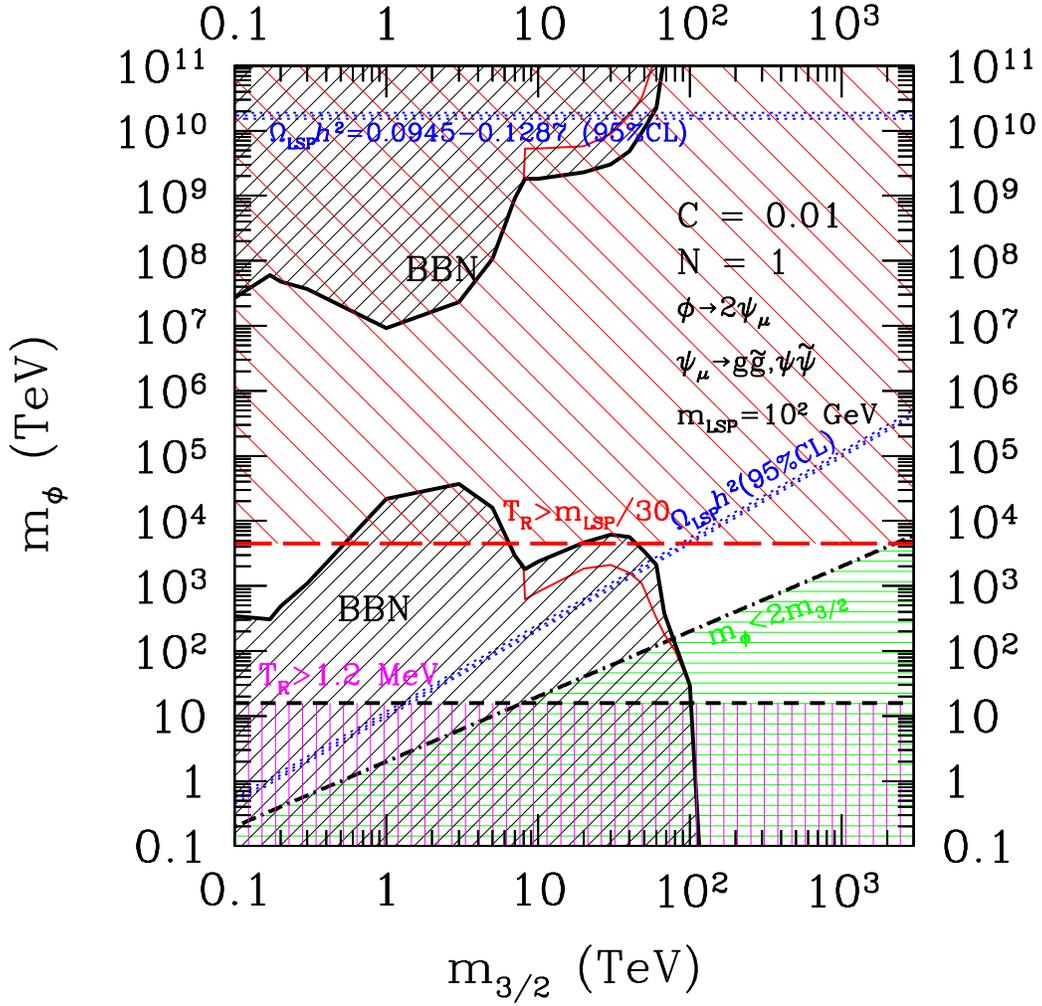}
\caption{ Various constraints on the $m_{3/2}$--$m_\phi$ plane, (i)~BBN
constraints (thick and thin solid lines), (ii)~$T_R < m_{\rm LSP}/30$
(long dashed line), (iii)~$T_R >1.2$ MeV (dashed line), and
(iv)~$m_{\phi} > 2 m_{3/2}$ (dot-dashed line) in the case $C=10^{-2}$
and $N=1$.  Also depicted there are two strips where neutralino
abundance from decaying gravitinos is in accord with the recently
observed value $\Omega_{\rm LSP}h^2=0.0945-0.1287~(95\%~{\rm C.L.})$
with $m_{\rm LSP} = 100$ GeV. There is a region on this strip where
all the constraints are satisfied simultaneously with $m_{3/2}\simeq
55-100$ TeV and $m_\phi\simeq (2-4)\times 10^3$ TeV. Since $N$ appears
only in the form $N^{1/2} \mphi^{3/2}$ in all of the relevant
expressions in
Eqs.~(\ref{eq:TR}),~(\ref{eq:thermal})~and~(\ref{eq:direct}), the
constraints for $N$ other than $N = 1$ can easily be read off by
replacing $\mphi$ in the vertical axis by $N^{1/3} \mphi$.}
\label{fig:two_grav_all_particle}
\end{center}
\end{figure}

We can now identify regions of the parameter space where the LSP
neutralino from the gravitino decay becomes the dark matter of the
universe.  Fig.~\ref{fig:two_grav_all_particle} depicts various
constraints on the $m_{3/2}$--$m_\phi$ plane, (i)~BBN constraints
(thick and thin solid lines), (ii)~$T_R < m_{\rm LSP}/30$ (long dashed
line), (iii)~$T_R >1.2$ MeV (dashed line), and (iv)~$m_{\phi} > 2
m_{3/2}$ (dot-dashed line) in the case $C=10^{-2}$ and $N=1$.  Also
depicted there are two strips where neutralino abundance from decaying
gravitinos is in accord with the recently observed value $\Omega_{\rm
LSP}h^2=0.0945-0.1287~(95\%~{\rm C.L.})$ with $m_{\rm LSP} = 100$ GeV.
The upper strip corresponds to the case gravitinos are thermally
produced after modulus decay.  However, it is entirely covered by the
region with $T_R >m_{\rm LSP}/30$ where the reheat temperature is so
high that LSPs are thermally populated after modulus decay.  In this
case dark matter LSPs tend to be overpopulated unless some enhancement
mechanism of annihilation is operative.  This long dashed line has a
scaling law for the other values of $m_{\rm LSP}$ as $m_{\phi} \propto
m_{\rm LSP}^{2/3}$.  The lower strip, on the other hand, represents
the case gravitinos are produced directly by modulus decay.  There is
a region on this strip where all the constraints are satisfied
simultaneously with $m_{3/2}\simeq 55-100$~TeV and $m_\phi\simeq
(2-4)\times 10^3$~TeV.  As is seen in the figure it is highly
nontrivial that such a region really exists. Since $N$ appears only in
the form $N^{1/2} \mphi^{3/2}$ in all of the relevant expressions in
Eqs.~(\ref{eq:TR}),~(\ref{eq:thermal})~and~(\ref{eq:direct}), the
constraints for $N$ other than $N = 1$ can easily be read off by
replacing $\mphi$ in the vertical axis by $N^{1/3} \mphi$.  Concerning
this lower strip, there is a scaling relation $\mphi \propto m_{\rm
LSP}^{2/3}$. Thus, for the other values of $m_{\rm LSP}$ we can easily
see how the results change by using this relation.

\begin{figure}[hp]
\begin{center}
\epsfxsize=0.9\textwidth\epsfbox{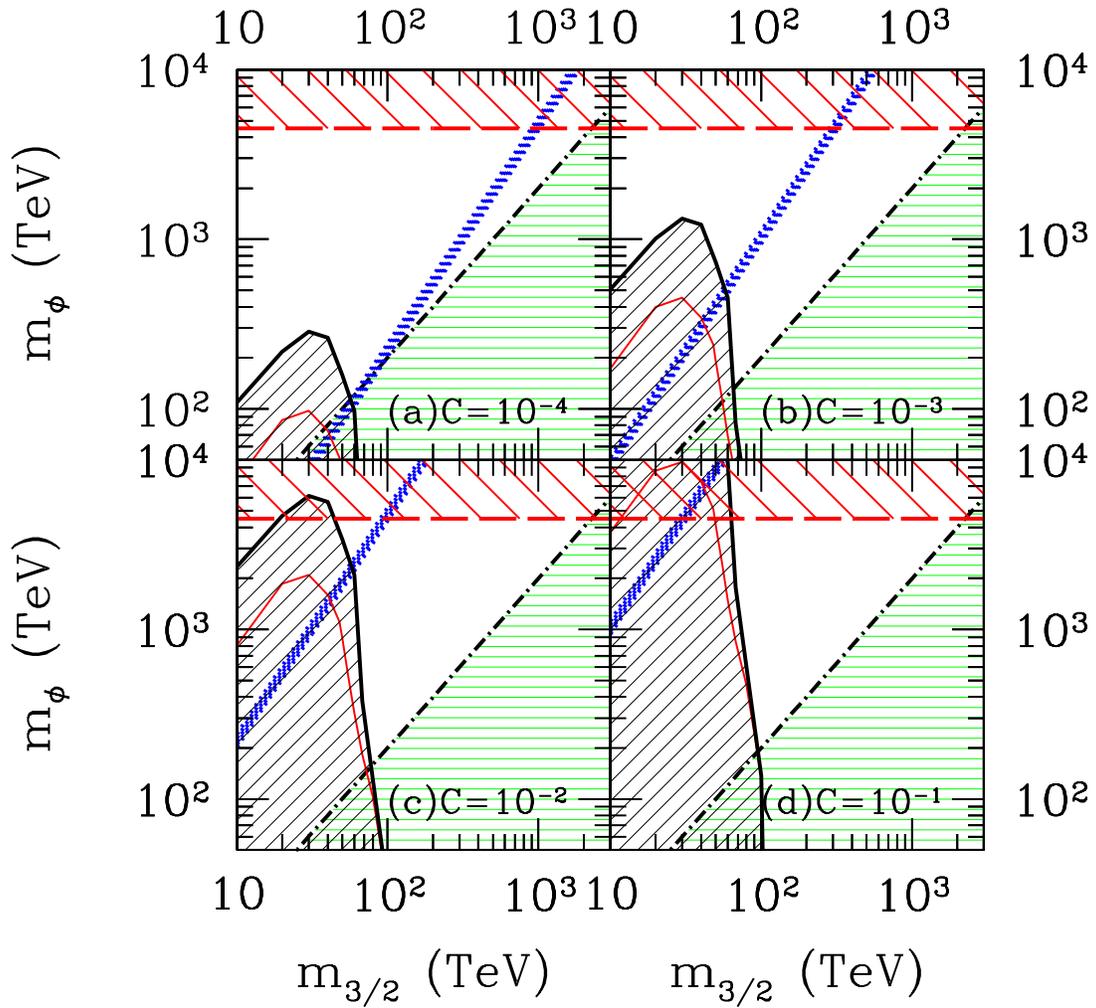}
\caption{How the results change as we take different values of $C$,
which parameterizes the branching ratio of the modulus decay into the
gravitino pair. We plot the cases (a) $C=10^{-4}$, (b) $C=10^{-3}$,
(c) $C=10^{-2}$, and (d) $C=10^{-1}$, respectively. For the
explanation of the various constraints, see the figure caption in
Fig.~\ref{fig:two_grav_all_particle}.}
\label{fig:panel_2}
\end{center}
\end{figure}

Fig.~\ref{fig:panel_2} depicts how the results change as we take
different values of $C$ which parameterizes the branching ratio of the
modulus decay into the gravitino pair.  We plot the cases (a)
$C=10^{-4}$, (b) $C=10^{-3}$, (c) $C=10^{-2}$, and (d) $C=10^{-1}$,
respectively.  From this panel, we can find that wide ranges of
$\mphi$ are allowed ($\sim$ $\order(10^{2})$ -- $\order(10^{3})$ TeV)
when $C$ is changed for $\sim \order(10^{-4})$ --
$\order(10^{-1})$. Here, note that the intersection between the trip
and the boundary of the BBN constraint gives a same value of $\mphi$
($\sim 55$ TeV) for any values of $C$.

\begin{figure}[hp]
\begin{center}
\epsfxsize=0.9\textwidth\epsfbox{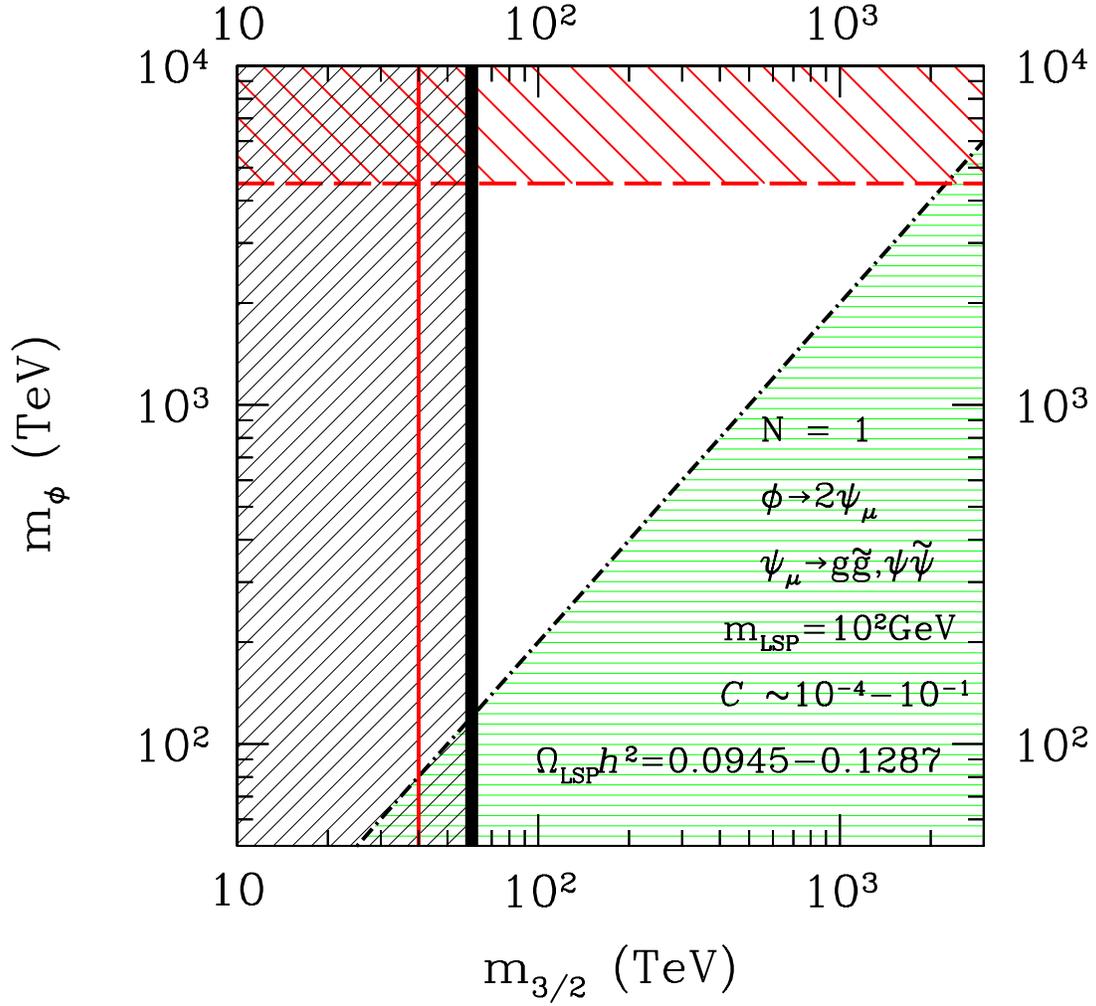}
\caption{Allowed region in the $m_{3/2}$--$m_\phi$ plane where
$\Omega_{\rm LSP}h^2$ can take the observed value with $C\sim
10^{-4}-10^{-1}$. The vertical thick (thin) solid line represents the
BBN constraint, which comes from the observational $^4$He mass
fraction by Izotov and Thuan $Y_p(IT)$ ( Olive and Skillman
$Y_p(OS)$). The long dashed line  represents the boundary of $T_R <
m_{\rm LSP}/30$ . This line for $N$ other than $N = 1$ can easily be
read off by replacing $\mphi$ in the vertical axis by $N^{1/3}
\mphi$. The dot-dashed line denotes the boundary of $m_{\phi} > 2
m_{3/2}$.}
\label{fig:window}
\end{center}
\end{figure}

In Fig.~\ref{fig:window} we depict the domain in the
$m_{3/2}$--$m_\phi$ plane where $\Omega_{\rm LSP}h^2$ can take the
observed value with some value of $C$ in the range $C\sim
10^{-4}-10^{-1}$.  The vertical thick (thin) solid line represents the
BBN constraint, which comes from the observational $^4$He mass
fraction by Izotov and Thuan $Y_p(IT)$ ( Olive and Skillman
$Y_p(OS)$). The long dashed line represents the boundary of $T_R <
m_{\rm LSP}/30$ . The dot-dashed line denotes the boundary of
$m_{\phi} > 2 m_{3/2}$.

As is seen there, we have a fairly large allowed
region  for $m_{3/2} \simeq$ 55 -- $2\times 10^{3}$~TeV and $m_{\phi}
\simeq 10^{2}$ -- $4\times 10^{3}$~TeV, with a small hierarchy between
$m_\phi$ and $m_{3/2}$.

For different values of $m_{\rm LSP}$, because the boundary of $T_R <
m_{\rm LSP}/30$ has a scaling law as $m_{\phi} \propto m_{\rm
LSP}^{2/3}$, we can rescale the long dashed line by using this
relation. In addition, about the modification of the vertical thick
and thin lines for the other values of $m_{\rm LSP}$, we can read off
their values of $m_{3/2}$ as the intersections between the trip and
the boundaries of the BBN constraint in Fig.~\ref{fig:m_n32_new} or
Fig.~\ref{fig:two_grav_all_particle}. The long dashed line for $N$
other than $N = 1$ can easily be read off by replacing $\mphi$ in the
vertical axis by $N^{1/3} \mphi$.

\section{Discussion}

 We have shown that the neutralino dark mater whose abundance
agrees with the WMAP constraint can be non-thermally produced by the
decay of heavy gravitino with the mass of $m_{3/2} \sim$ 55 -- 100 TeV
in the early universe. Such gravitinos are produced only by the
decaying moduli with the late-time large entropy production, which
completely dilutes both the thermal relics of LSPs, and primordial
gravitinos produced by the reheating process after the primordial
inflation. In this case, we found that the required region of the
moduli mass is $m_{\phi} \simeq 10^{2}$ -- $4\times 10^{3}$~TeV. This
scenario can solve the overproduction problem of LSPs in the
supersymmetric cosmology. 

Note that although the LSP is still the best candidate for cold dark
matter in particle physics, the recent measurement of the abundance by
the WMAP restricts the preferred regions of the SUSY parameter
space. The old and naive argument that the dark matter abundance is of
the order of the critical density of the universe, which allows the
wide regions of the parameter space, does no longer apply today
because its abundance has been measured to be only one quarter of the
critical density, which turned out to be significantly smaller than we
expected in the standard calculation. As mentioned in the
introduction, attempts do exist to explain this smallness within the
standard thermal history, but this in turn requires us to adopt some
enhancement mechanism of pair annihilation such as coannihilation or
resonance enhancement etc (For the details see, {\em e.g.}   recent
works~\cite{Ellis:2004qe,Roszkowski:2004jd}).  Our scenario is in a
sense more economical than these mechanisms because we can account for
the desired abundance of the neutralino dark matter as these particles
are produced from modulus decay which also produces large entropy to
solve the new gravitino problem at the same time.

It is interesting to point out an alternative application of the
production mechanism described in this paper. The neutralino
production by gravitino decay can allow the neutralino dark matter
with large fraction of the Wino and Higgsino components: its thermal
abundance would be too small to constitute the dark matter of the
universe. In this case, the cross section of the neutralino-nuclei
scattering and the neutralino-neutralino pair annihilation can become
larger. Thus, in the direct searches such as CDMS~\cite{ref:CDMS},
EDELWEISS~\cite{ref:edelweiss}, and DAMA~\cite{ref:DAMA}, we expect a
higher detection rate than that of pure Bino LSP. In the indirect
searches, we can also expect higher detectability of the signal of
photons, neutrinos, electrons and their radio emission through the
neutralino-neutralino pair annihilation in the galactic center in the
Milky Way, the center of the other galaxies, and the clusters of
galaxies in the future experiments such as GLAST~\cite{ref:GLAST} and
MAGIC~\cite{ref:MAGIC} for gamma-rays, and
Hyper-Kamiokande~\cite{ref:hyper-K} and TITAND~\cite{Suzuki:2001rb}
for neutrinos, and so on.

Finally we would like to emphasize that in this scenario the large
gravitino mass will make the superparticle mass spectrum  very
different from the conventional gravity mediated SUSY breaking, and we
expect the admixture of gravity mediation and anomaly
mediation~\cite{Choi:2005ge,Endo:2005uy,Choi:2005uz,Falkowski:2005ck},
which will be testable in future collider experiments.

\section*{Acknowledgments}

This work  was partially  supported by the  JSPS Grants-in Aid  of the
Ministry  of   Education,  Science,  Sports,  and   Culture  of  Japan
No.~15-03605 (KK), No.~13640285 and No.~16340076 (JY), and No.12047201
(MY). K.K. was also supported by NSF grant AST 0307433.

\newpage



\end{document}